\documentclass[conference]{IEEEtran}
\IEEEoverridecommandlockouts
\usepackage{cite}
\usepackage{amsthm, amsmath,amssymb,amsfonts}
\usepackage{algorithmic}
\usepackage{graphicx}
\usepackage{textcomp}
\usepackage{multirow}
\usepackage{array}
\usepackage{hhline}
\usepackage{xcolor}
\usepackage{hyperref}
\usepackage{kotex}
\usepackage{makecell}
\usepackage{comment}
\usepackage{tablefootnote}

\newtheorem{theorem}{Theorem}[section]

\newcommand\T{\rule{0pt}{2.6ex}}       %
\newcommand\B{\rule[-1.2ex]{0pt}{0pt}} %

\newcolumntype{x}[1]{>{\centering\arraybackslash\hspace{0pt}}p{#1}}

\def\BibTeX{{\rm B\kern-.05em{\sc i\kern-.025em b}\kern-.08em
    T\kern-.1667em\lower.7ex\hbox{E}\kern-.125emX}}
    
\begin{document}

\title{Do You Really Need to Disguise Normal Servers \\
as Honeypots?
}

\author{\IEEEauthorblockN{Suhyeon Lee\IEEEauthorrefmark{1}\IEEEauthorrefmark{2}, Kwangsoo Cho\IEEEauthorrefmark{2}, and Seungjoo Kim\IEEEauthorrefmark{2}}
\IEEEauthorblockA{\IEEEauthorrefmark{1}Cyber Operations Command, Republic of Korea}
\IEEEauthorblockA{\IEEEauthorrefmark{2}School of Cybersecurity, Korea University \\
Email: \{orion-alpha, cks4386, skim71\}@korea.ac.kr}}
\maketitle

\begin{abstract}
A honeypot, which is a kind of deception strategy, has been widely used for at least 20 years to mitigate cyber threats. Decision-makers have believed that honeypot strategies are intuitive and effective, since honeypots have successfully protected systems from Denial-of-Service (DoS) attacks to Advanced Persistent Threats (APT) in real-world cases.
Nonetheless, there is a lack of research on the appropriate level of honeypot technique application to choose real-world operations. We examine and contrast three attack-defense games with respect to honeypot detection techniques in this paper. In particular, we specifically design and contrast two stages of honeypot technology one by one, starting with a game without deception.
We demonstrate that the return for a defender using honeypots is higher than for a defender without them, albeit the defender may not always benefit financially from using more honeypot deception strategies. Particularly, disguising regular servers as honeypots does not provide defenders with a better reward. Furthermore, we take in consideration that fake honeypots can make maintaining normal nodes more costly.
Our research offers a theoretical foundation for the real-world operator's decision of honeypot deception tactics and the required number of honeypot nodes.

\end{abstract}


\section{Introduction}

Cyber attacks are getting more threatening as a consequence of the proliferation of digital technologies such as cloud computing and the Internet of Things (IoT). Defenders create strategies to counterattack. Accordingly, attackers are persistent in developing new methods.
A honeypot is one technology that enables cybersecurity agents to trap attackers and collect threat intelligence. This intelligence ultimately enables them to learn and strengthen safeguards against future threats. However, only establishing a large number of honeypots is not a viable option, and installing and operating honeypots requires a strategic approach.

Cybersecurity can utilize game theory to analyze the most effective techniques \cite{do2017game, pawlick2019game}. Game theory has applications in all social science disciplines, as well as logic, systems science, and computer science. Originally, it addressed zero-sum games, in which each player's earnings or losses are exactly balanced by those of the other players.
In game-theoretic examination of honeypot technology, it can be described as a signaling game in which the defender indicates whether a specific node is honeypot or normal. These models were the subject of a significant investigation in \cite{carroll2011game}. Perfect Bayesian Equilibrium (PBE) was investigated using a signaling game with symmetric payoffs. They derived 10 equilibria that every node sends the same honeypot or normal signal.

La et al. \cite{la2016deceptive} analyzed honeypot defense strategies in Internet of Things (IoT). In their model, an attacker sends a signal and the defender chooses the defense strategy according to the signal. Li et al. \cite{li2020anti} analyzed signaling games with anti-honeypot techniques in industrial systems. Diamantoulakis et al. \cite{diamantoulakis2020game} studied the optimal honeypot ratio by analyzing the strategy of switching nodes to honeypot in an environment where no new nodes are added.
Nevertheless, systems must retain their normal nodes to maintain service quality. In this perspective, we focus on the number of honeypot nodes rather than the number of defensive nodes.
Shortridge \cite{2017Shortridge} claimed that making defenders' environment resemble an analyst's sandbox can be a good strategy from a practical perspective.
We found her reasoning to be really compelling. This study concentrated on how the payoffs of defenders vary as honeypot nodes and fake honeypot nodes are gradually added.

Our contributions are as follows:

    \begin{itemize}
        \item We show that profits do not always increase even if the number of defense techniques increase in the honeypot game. This research applies zero to three honeypot deception actions to attacker-defender games. In the presence of additional cost in the normal node deception, we conclude that deception techniques for normal nodes are practically ineffective in choosing the best strategy.
        
        \item We demonstrate that an increase in the number of honeypot nodes does not always increase the payoff of the defender. The defender's payoff continues to increase to a certain point, but continues to decrease beyond the maximum point. Furthermore, we confirm that payoff can be dramatically reduced assuming that the honeypot cost is dynamic.
    \end{itemize}

The paper is organized as follows.
In Section \ref{section: background}, we give an overview of the background of the honeypot deception and the signaling game.
In Section \ref{section: model}, we describe a scenario and models of honeypot deception games.
In Section \ref{section: analysis}, we analyze the equilibria of the signaling games of honeypot deception.
In Section \ref{section: optimal decision}, we find optimal honeypot distributions based on the analysis. Finally, we examine the dynamic payoff in the honeypot deception game.
In Section \ref{section: case study}, we show cases of signaling games with a fixed cost and a dynamic cost of honeypot nodes.
In Section \ref{section: conclusions}, conclusions are presented.

    \begin{figure}[b!]
        \includegraphics[width=\linewidth]{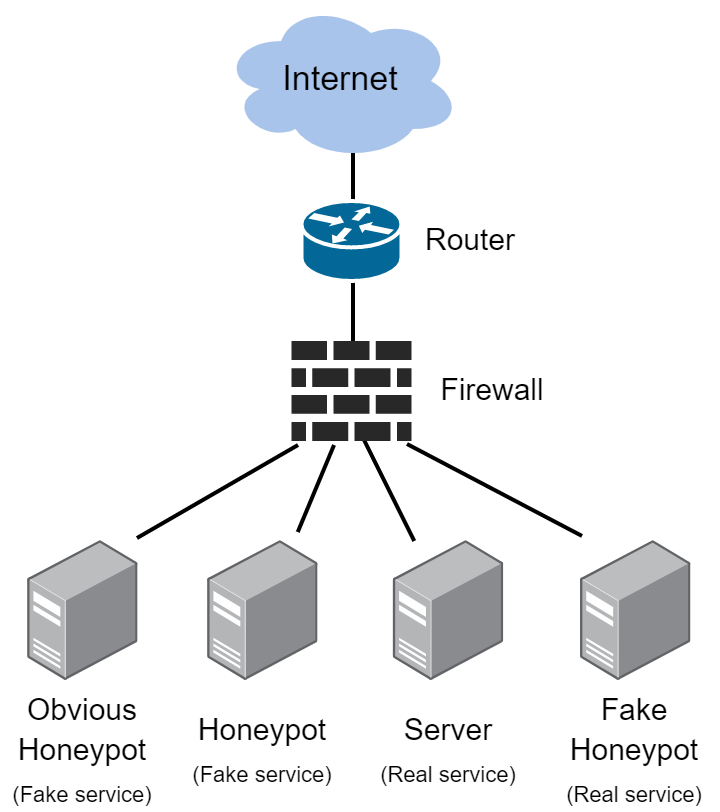}
        \centering
        \caption{Honeypot Deception Game}
        \label{fig: game overview}
    \end{figure}

\section{Background} \label{section: background}

    In this section, we provide an overview of honeypot deception strategies and a signaling game in a game theory.

    \subsection{Honeypot Deception}
    
    The honeypot technology is a methodology of avoiding attacks or analyzing attacks by attracting attackers. Honeypot techniques can be used to defend not only low level-attacker but also high-level attacks related to industrial control systems \cite{li2020anti, la2016deceptive, disso2013plausible}, DoS \cite{kramer2015amppot}, and APT \cite{jasek2013apt}. Fig. \ref{fig: game overview} illustrates basic elements of honeypot options. Conceptually, in a honeypot strategy, honeypots should look like a normal system for attackers. In Fig. \ref{fig: game overview}, the second server `honeypot' is such a concept. Since attackers try to avoid honeypot systems, attempts have emerged to disguise a normal node as a honeypot node. It is called normal-as-honeypot which is the fourth server in Fig. \ref{fig: game overview}. 
    
    From the attacker's point of view, attackers need anti-honeypot techniques. Conversely, the defender's technique to prevent attackers' investigation is termed an anti-introspection technique \cite{uitto2017survey, krawetz2004anti}. It is the crux of the honeypot strategy that attackers and defenders deceive and avoid each other persistently. Depending on the strategic situation, it may also be useful to be clearly seen as a honeypot node. For example, if it is determined that a defender disguises a normal server as a honeypot, the attacker targets a server that has features like a honeypot. It is the first server in Fig. \ref{fig: game overview}

    \subsection{Signaling Game}
    
    A signaling game is a two-player incomplete information game. A sender and a receiver participate in the game. The sender has private information. The sender is selected under a certain probability as one of types that are provided by the game model. Then the sender selects a signal type and sends the signal (or message) to the receiver. The receiver observes the signal and selects an action. At last, their payoffs are decided. The equilibrium in a signaling game is Perfect Bayesian Equilibrium (PBE), a refined concept of Bayesian Nash Equilibrium (BNE). In a Bayesian game based on incomplete information, players have types and beliefs. Types are decided by a special player called `nature' for convenience. Beliefs are the probability distribution of one player's signal and type by another player. In PBE at least, each player's strategy should be a best response in the given beliefs. At the same time, each player's strategy should be a best response to the updated belief. For off-equilibrium paths, the beliefs can be arbitrary. However, some arbitrary beliefs can be irrational so that a PBE which relies on such beliefs can be eliminated with advanced refinement rules \cite{cho1987signaling}.

    \begin{figure*}[tb!]
        \includegraphics[width=\linewidth]{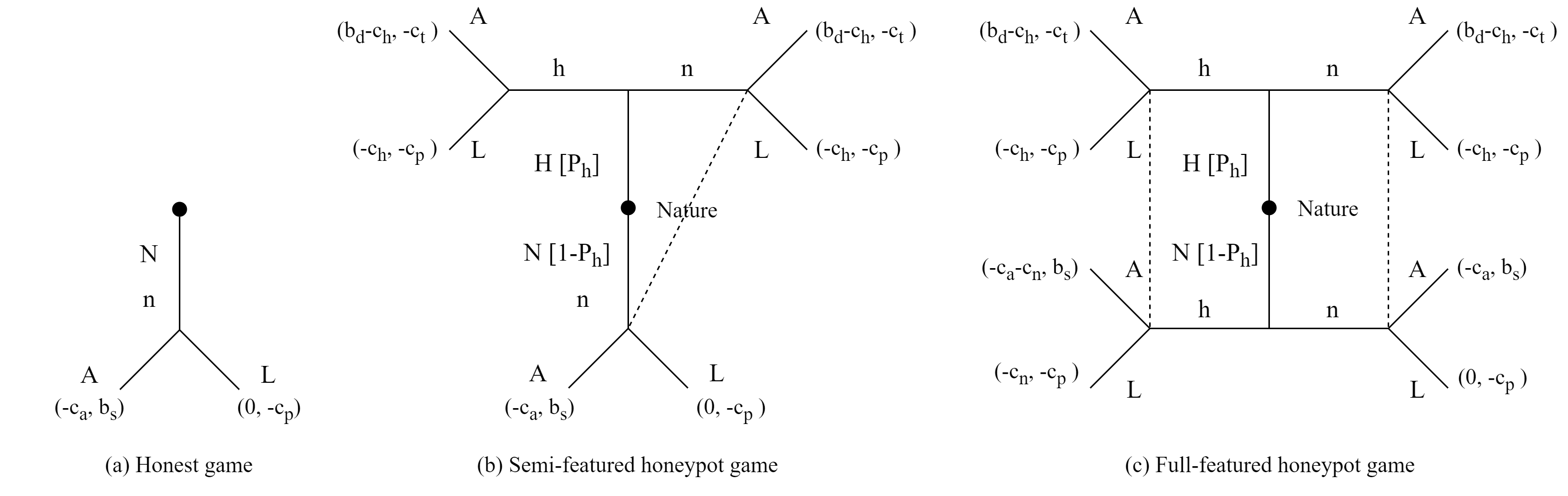}
        \centering
        \caption{Honeypot Deception Game Models}
        \label{fig: game model}
    \end{figure*}

\section{Honeypot Deception Game Model} \label{section: model}

    In this section, we describe the scenario and models of the honeypot deception game. Table \ref{table: List of symbols} shows the notation associated with the game models. Table \ref{table: List of payoff} describes the payoffs of attackers and defenders.

    \begin{table}[b!]
    \centering
    \caption{List of notations}
    \label{table: List of symbols}
    \begin{tabular}{lp{0.37\textwidth}}
    \Xhline{2\arrayrulewidth}
    Notation & Description      \T\B                                                                       \\ \hline %
    H       &   A node is honeypot  \\ \hline
    N       &   A node is normal    \\ \hline
    h       &   Signal that the node is honeypot    \\ \hline
    n       &   Signal that the node is normal      \\ \hline
    A       &   Attack the node     \\  \hline
    L       &   Leave the node without attack   \\  \hline
    $P_h$     & Probability that nature selects a node as honeypot                                  \\ \hline
    $1-P_h$   & Probability that nature selects a node as normal                                    \\ \hline
    p      & Probability that the node observed as normal to the attacker is actually honeypot     \\ \hline
    1-p    & Probability that the node observed as normal to the attacker is actually normal   \\ \hline
    q      & Probability that the node observed as honeypot to the attacker is actually honeypot   \\ \hline
    1-q    & Probability that the node observed as honeypot to the attacker is actually normal \\ \Xhline{2\arrayrulewidth}
    \end{tabular}
    \end{table}

    \begin{table}[tb!]
    \centering
    \caption{List of payoff}
    \label{table: List of payoff}
    \begin{tabular}{cx{1.3cm}p{5.2cm}}
    \Xhline{2\arrayrulewidth}
        &   Parameter   &   Description  \T\B   \\    \hline
    \multirow{3}{*}{Attacker} & $b_s$ & Attacker's benefit from compromising the system           \\ %
                              & $c_t$ & Attacker's cost from revealing attack on honeypot \\ %
                              & $c_p$ & Attacker's cost from probing honeypot             \\ \hline
    \multicolumn{3}{c}{}                                                                 \\ \hline
    \multirow{4}{*}{Defender} & $b_d$ & Defender's benefit from detection of attack        \\ %
                              & $c_a$ & Defender's cost from successful attack            \\ %
                              & $c_h$ & Defender's cost from honeypot node deployment          \\ %
                              & $c_n$ & Defender's cost from normal-as-honeypot node deployment          \\ \Xhline{2\arrayrulewidth}
    \end{tabular}
    \end{table}

    \subsection{The attacker-defender honeypot game scenario}
    
    The attacker-defender honeypot game scenario fundamentally consists of attackers outside invading a defender's network as illustrated in Fig. \ref{fig: game overview}. The attacker finds a node in the defender's network and decides whether to attack it or leave without attacking it. The defender's network consists of two types: a normal node and a honeypot node. When the attacker attacks a normal node, they benefit from achieving the desired goal, and the defender receives damage. When the attacker attacks a honeypot node, the attack method is exposed, allowing the defender to counter the attack method in advance. This causes the attacker to get damage and the defender to profit. The attacker can observe whether it is a normal node or a honeypot node depending on the signal sent by the defender as we assume the defender uses decent deception techniques. If the attacker leaves without attacking a node, the defender does not get harmed. Then, the attacker wastes time without any profit. The defender has to pay as much as the number of honeypot nodes. The ratio of the defender's honeypot nodes is determined through factors such as budget and policy.

    \subsection{The attacker-defender Game Model}

    The game models present three attacker-defender games. Fig. \ref{fig: game model} (a) represents the first game. The first game is in a state where defenders do not use any honeypot techniques. This game is named an honest game. Only normal nodes are given, and the attacker decides whether to attack or leave. When the attacker targets a normal node, the defender suffers a loss ($-c_a$) due to the attack. On the other hand, if the attacker just leaves without attacking the target, they spend time navigating the node, resulting in a loss ($-c_p$). The defender experiences neither profit or loss.

    Figure \ref{fig: game model} (b) illustrates the second game. In the second game, the defender runs a honeypot node. The defender is assigned a honeypot node from nature with a probability ($P_h$), where a defender can choose between a honeypot node signal (h) and a normal node signal (n). The attacker observes a honeypot node signal (h) and a normal node signal (n). The attacker cannot distinguish between a normal node and a honeypot node through a signal (n). This game is named a semi-featured honeypot game. Even though the next game model has more options for the defender, this semi-featured honeypot game is realistic in many organizations as they do not own normal-as-honeypot techniques or plan to apply it yet.
    If the attacker decides to attack a node which is actually a honeypot regardless of the type of signal, the attacker suffers a loss ($-c_t$) because it exposes the attack method. The defender gains a profit ($b_d$) by detecting the attack. On the other hand, the defender costs ($-c_h$) as the defender deployed honeypot. If the attacker decides not to attack a honeypot node, the attacker suffers a loss ($-c_p$) because it has spent time navigating that node. We suppose that $c_t > c_p$ as revealing concrete attack methods will be more costly than simply proving nodes.
    For a normal node (N) that sent a normal node signal (n), the attacker gains  a profit ($b_s$) from a successful attack. The defender pays $-c_a$ due to a successful attack. If the attacker avoids a normal node, the defender has a 0 payoff. The attacker suffers a loss ($-c_p$) because it has spent time navigating that node. Also, it is assumed that $c_h < c_a$ as the cost by successful attacks will be the most threatening.
    
    Figure \ref{fig: game model} (c) illustrates the third game. This game is named a full-featured honeypot game. Like the second game, the defender runs a honeypot node. The defender is assigned a honeypot node from nature with a probability ($P_h$), where a defender can choose between a honeypot node signal (h) and a normal node signal (n). With a probability ($1-P_h$), a normal node is assigned by nature, where the defender can send a honeypot node signal (h) or a normal node signal (n). The attacker observes a honeypot node signal (h) and a normal node signal (n). The attacker cannot distinguish between a normal node and a honeypot node through a signal (n) or a signal (h).
    If the attacker decides to target a node that is actually a honeypot regardless of the type of signal, the attacker suffers a loss ($-c_t$) because it exposes the attack method. The defender gains a profit ($b_d$) by detecting the attack. On the other hand, the defender pays ($-c_h$) as the defender deployed honeypot. If the attacker decides not to attack a honeypot node, the attacker suffers a loss ($-c_p$) because it has spent time navigating that node. We suppose that $c_t > c_p$.
    If the attacker decides to attack or leave a node, it goes to the same with the second model. On the other hand, the defender costs ($-c_n$) as the defender deployed a normal-as-honeypot node. It is assumed that $c_n < c_h$. It is because the cost to deploy a normal-as-honeypot node can be handled in a software method to an existing normal node. Also, it is assumed that $c_n < c_a$ and $c_h < c_a$ as the cost by successful attacks will be the most threatening.

\section{Analysis} \label{section: analysis}

    In this section, we analyze equilibrium strategies of the three games introduced in Section \ref{section: model}.

    \subsection{Equilibria Analysis on Honest Game}
    
    Obviously, in this game, HE1 in Table \ref{table: Equilibrium of honest game} is the only equilibrium point because the defender does not have a choice for the signal, and the attacker's dominant strategy is to attack the normal node. At this point, the payoff is ($-c_a, b_s$).

    \begin{table}[!b]
    \centering
    \caption{Equilibrium of honest game}
    \label{table: Equilibrium of honest game}
    \begin{tabular}{x{1.5cm}x{3cm}x{3cm}}
    \Xhline{2\arrayrulewidth}
        & Strategy & Conditions \T\B \\ \hline
    HE1 & (A)      & -          \T\B \\ \Xhline{2\arrayrulewidth}
    \end{tabular}
    \end{table}
    
    \subsection{Equilibria Analysis on Semi-featured Honeypot Game}
    
    Let (a, (B, C)) denote a strategy set where `a' is the H-type defender's action, `B' is the attacker's action when the signal (h) is received, and `C' is the attacker's action when the signal (n) is received.
    In this game, (n, (L, A)) and (n, (L, L)) are the only balance points, as shown in Table \ref{table: equilibria of semi featured honeypot game}. When the defender's type is H, if the defender sends an h signal, the attacker chooses an attack strategy (L) because it can clearly tell if it is a honeypot. Therefore, sending the signal (n) becomes a dominant strategy. If the defender always sends the signal (n), the honeypot and normal nodes cannot be distinguished, thus determining the strategy relative to the size of $P_h$. To choose the strategy (A), the attacker's payoff must be greater than the strategy L's payoff when choosing the strategy (A). Therefore, $P_h$ must satisfy the following:
    
    \begin{equation}
        P_h \cdot (-c_t) + (1-P_h) \cdot b_s \geq P_h \cdot (-c_p) + (1 - P_h) \cdot (-c_p).
    \end{equation}
    
    \noindent In summary, the following inequality is obtained:
    
    \begin{equation}
        P_h \leq \frac{b_s + c_p}{b_s + c_t}
    \end{equation}
    
    In this case, the strategy for the signal (h) becomes an off-equilibrium path, but it is apparent that for the signal (h), regardless of the defender's belief, the strategy L becomes the dominant strategy.

    \subsection{Equilibria Analysis on Full-featured Honeypot Game}
    
    Let ((a, b), (C, D)) denote a strategy set where `a' is the H-type defender's action, `b' is the N-type defender's action, `C' is the attacker's action when the signal (h) is received, and `D' is the attacker's action when the signal (n) is received. First, separating equilibrium does not exist in this game. If the defender attempts a separating strategy, the signal allows the attacker to pinpoint the node's type without uncertainty. In this instance, the attacker uses an attack strategy (A) for normal nodes and an avoidance strategy (L) for honeypot nodes. In this state, it is a dominant strategy for the defender to reverse the signal.

    Second, two pooling equilibria exist in this game. Table \ref{table: equilibria of full featured honeypot game} shows two pooling equilibria under the scenario that the defender chooses (n, n). Below, we illustrate how FE1 and FE2 satisfy the PBE conditions and why the (h, h) strategy cannot have a PBE.

    We look at FE1's strategies ((n, n), (A, A)). The payoff obtained by the defender modifying the (n, n) strategy to the (h, n) strategy is the same. The payoff of the normal node obtained by the defender modifying the (n, n) strategy to the (n, h) strategy is $-c_a - c_n$, which is smaller than the existing payoff, $c_a$. Therefore, the defender has no incentive to change the strategy. For the attacker, in order to adhere to the strategy A for the signal n, the following must be satisfied:

    \begin{equation}
        P_h (-c_t) + (1-P_h) b_s \geq P_h (-c_p) + (1 - P_h) (-c_p).
    \end{equation}
    
    \noindent Like SE1, it gives the following inequality:
    
    \begin{equation}
        P_h \leq \frac{b_s + c_p}{b_s + c_t}.
    \end{equation}
    
    \noindent To adhere to the strategy A for the signal h, the following belief must be supported for the off-equilibrium path:

    \begin{equation}
        q (-c_t) + (1-q) b_s \geq q (-c_p) + (1 - q) (-c_p)
    \end{equation}
    
    \noindent which results in
    
    \begin{equation}
        q \leq \frac{b_s + c_p}{b_s + c_t}.
    \end{equation}
    
    \noindent For the conditions of FE2, we go through the same process as FE1 to show that the opposite conditions are constraints.

    Next, we look at the (h, h) strategy. For the defender to send the signal (h) for the type (n), it costs $-c_n$. ((h, h), (A, A)) strategy results in $-c_a - c_n < -c_a$ for the normal node when the defender deviates to (h, n). In this case, the defender is motivated to change the strategy. ((h, h), (L, L)) strategy gives $c_n < 0$ for the normal node when the defender changed to (h, n). Then, the defender has an incentive to change the strategy. Likewise, (h, h), (L, A)) and ((h, h), (A, L)) strategies can be deviated for better payoffs.

    \begin{table}[!tb]
    \centering
    \caption{Equilibria of semi-featured honeypot game}
    \label{table: equilibria of semi featured honeypot game}
    \begin{tabular}{x{1cm}x{1.7cm}x{2cm}x{2cm}}
    \Xhline{2\arrayrulewidth}
    \multirow{2}{*}{} & \multirow{2}{*}{PBE} & \multicolumn{2}{c}{Conditions}                                         \T\B  \\ \cline{3-4} 
                      &                      & \multicolumn{1}{c}{On-equilibrium} & \multicolumn{1}{c}{Off-equilibrium} \T\B \\ \hline
    SE1               & (n, (L,A))           & $P_h \leq \frac{b_s + c_p}{b_s + c_t}$                               & -                              \T\B  \\ \hline
    SE2               & (n, (L,L))           & $P_h \geq \frac{b_s + c_p}{b_s + c_t}$                               & -                              \T\B  \\ \Xhline{2\arrayrulewidth}

    \end{tabular}
    \end{table}

    \begin{table}[!tb]
    \centering
    \caption{Equilibria of full-featured honeypot game}
    \label{table: equilibria of full featured honeypot game}
    \begin{tabular}{x{1cm}x{1.7cm}x{2cm}x{2cm}}
    \Xhline{2\arrayrulewidth}
    \multirow{2}{*}{} & \multirow{2}{*}{PBE} & \multicolumn{2}{c}{Conditions}                                         \T\B  \\ \cline{3-4} 
                      &                      & \multicolumn{1}{c}{On-equilibrium} & \multicolumn{1}{c}{Off-equilibrium} \T\B \\ \hline
    FE1               & ((n,n), (A,A))       & $P_h \leq \frac{b_s + c_p}{b_s + c_t}$                               & $q \leq \frac{b_s + c_p}{b_s + c_t}$                             \T\B   \\ \hline
    FE2               & ((n,n), (L,L))       & $P_h \geq \frac{b_s + c_p}{b_s + c_t}$                               & $q \geq \frac{b_s + c_p}{b_s + c_t}$                             \T\B   \\ \hline
    FE3               & ($\tau$, $\sigma$)\tablefootnote{See Theorem \ref{theorem: semi-separating PBE}}       & $P_h = \frac{b_s + c_p}{b_s + c_t}$                               & -                         \T\B   \\
    \Xhline{2\arrayrulewidth}
    \end{tabular}
    \end{table}

\begin{theorem} \label{theorem: semi-separating PBE}

The propose honeypot deception game (c) in Fig. \ref{fig: game model} has a semi-separating 
PBE $(\tau, \sigma)$ where,
\begin{equation}
    \begin{cases}
        \tau = & \big( (\alpha, 1-\alpha), (\alpha, 1-\alpha) \big) \text{ where $\alpha \in (0, 1)$,} \\
        \sigma = & \big( (\frac{c_h}{b_d + c_a}, \frac{b_d + c_a - c_h}{b_d + c_a}), (\frac{c_h}{b_d + c_a}\text{,} \frac{b_d + c_a - c_h}{b_d + c_a}) \big), \\
        \text{with the belief} & p = \frac{b_s + c_p}{b_s + c_t}, q= \frac{b_s + c_p}{b_s + c_t}, \\
        \text{when} & P_h = \frac{b_s + c_p}{b_s + c_t}.
    \end{cases}
\end{equation}

\end{theorem}

\begin{proof}

Let there exist a semi-separating PBE with an attacker's strategy $\big( (\sigma_a, 1 - \sigma_a), (\sigma_b, 1 - \sigma_b) \big)$ with a belief $(p, q)$.

Applying the indifference principle, the utilities of the defender's actions given the honeypot type should be same.

\begin{equation} \label{equation: indifference 1}
    \sigma_a ( b_d - c_h ) + (1 - \sigma_a) (-c_h) = \sigma_b ( b_d - c_h) + (1 - \sigma_b) (-c_h)
\end{equation}

In the same way, the utilities of the defender's actions given the honeypot type should be same.

\begin{equation} \label{equation: indifference 2}
    \sigma_a ( -c_a - c_h) + (1 - \sigma_a) (-c_h) = \sigma_b (-c_a) + (1 - \sigma_b) \cdot 0
\end{equation}

Using Equation \ref{equation: indifference 1} and \ref{equation: indifference 2}, we get

\begin{equation}
    \sigma_a = \frac{c_h}{b_d + c_a} \text{, } \sigma_b = \frac{c_h}{b_d + c_a}.
\end{equation}

On the other hand, we can use the indifference principle on the utilities of the attacker's actions given the honeypot signal.

\begin{equation} \label{equation: indifference 3}
    p (-c_t) + (1 - p) b_s = p (-c_p) + (1 - p) (-c_p)
\end{equation}

Likewise, we can apply the principle to the normal signal.

\begin{equation} \label{equation: indifference 4}
    q (-c_t) + (1 - q) b_s = q (-c_p) + (1 - q) (-c_p)
\end{equation}

Using Equation \ref{equation: indifference 3} and \ref{equation: indifference 4}, we get

\begin{equation}
    p = \frac{b_s + c_p}{b_s + c_t} \text{, } q = \frac{b_s + c_p}{b_s + c_t}.
\end{equation}

Only $P_h = \frac{b_s + c_p}{b_s + c_t}$ can satisfy the belief $(p, q)$.

\end{proof}

\section{Deciding Optimal Honeypot Deployment}
\label{section: optimal decision}

This section reviews the honeypot deception game in the real world and finds optimal decisions for the deployment of honeypot deception strategies.

In the signaling game models of Section \ref{section: model}, the distribution of honeypot and normal nodes are given by nature. We derived multiple PBE in the given distribution. However, in practical use, we can control the distribution of honeypot nodes when we deploy the honeypot deception strategy to an existing system. For example, a system administrator decides to use honeypots to protect a system connected to networks. This problem can be interpreted as a problem of how many honeypot nodes the administrator deploys compared to normal nodes.
Therefore, we need to find optimal honeypot distributions ($P_h$ in the game models) for the decision.

For the defender's payoff, we can derive the following theorem.

\begin{theorem}
In the honeypot deception games (b) -- (c) in Fig. \ref{fig: game model}, the defender's payoff is maximized at the point $P_h  = \frac{b_s + c_p}{b_s + c_t}$.
\end{theorem}

\begin{proof}
Let the payoff of the defender be $D(P_h)$.
In $P_h \in [0, \frac{b_s + c_p}{b_s + c_t})$, $\frac{dD(P_h)}{dP_h} > 0$ as we assumed $c_a > c_h$. In $P_h \in (\frac{b_s + c_p}{b_s + c_t}, 1]$, $\frac{dD(P_h)}{dP_h} < 0$ . Therefore, $D(P_h)$ has a maximum at the point $\frac{b_s + c_p}{b_s + c_t}$ at least with one strategy.
\end{proof}

On the other hand, for the attacker's payoff, we can derive the following theorem.

\begin{theorem}
In the honeypot deception games (b) -- (c) in Fig. \ref{fig: game model}, the attacker's payoff with bigger $P_h$ is less than or equal to the attacker's payoff with smaller $P_h$.
\end{theorem}

\begin{proof}
Let the payoff of the defender be $A(P_h)$.
In $P_h \in [0, \frac{b_s + c_p}{b_s + c_t})$, $\frac{dA(P_h)}{dP_h} < 0$. In $P_h \in (\frac{b_s + c_p}{b_s + c_t}, 1]$, $D(P_h) = -c_h$.
\end{proof}

We can see that to control $P_h$ near $\frac{b_s + c_p}{b_s + c_t}$ gives a better payoff if possible because of the previous theorems. 
System resources, non-technical problems such as budgets, and even policies can make it difficult to achieve optimal honeypot distribution.
Besides, the previous theorems are based on the assumption that payoffs are constant. We have to consider that the payoff can change dynamically by the $P_h$ distribution. For example, the cost to deploy honeypot nodes $c_h$ increases as the number of honeypot nodes increases. 
Even if honeypot nodes are virtually deployed, we can reasonably assume that resource consumption is proportional to the number of exposure. We assume that normal nodes in the system meant to protect are considered fixed. Then, with the ratio of honeypot nodes is $P_h$, the number of honeypot nodes to normal nodes is $P_h / (1 - P_h)$. The camouflage cost $c_n$ of the normal node is constant because the number of normal nodes is constant, and only the honeypot node deployment cost $c_h$ changes in proportion to $P_h / (1 - P_h)$.

    \begin{figure}[tb!]
        \includegraphics[width=\linewidth]{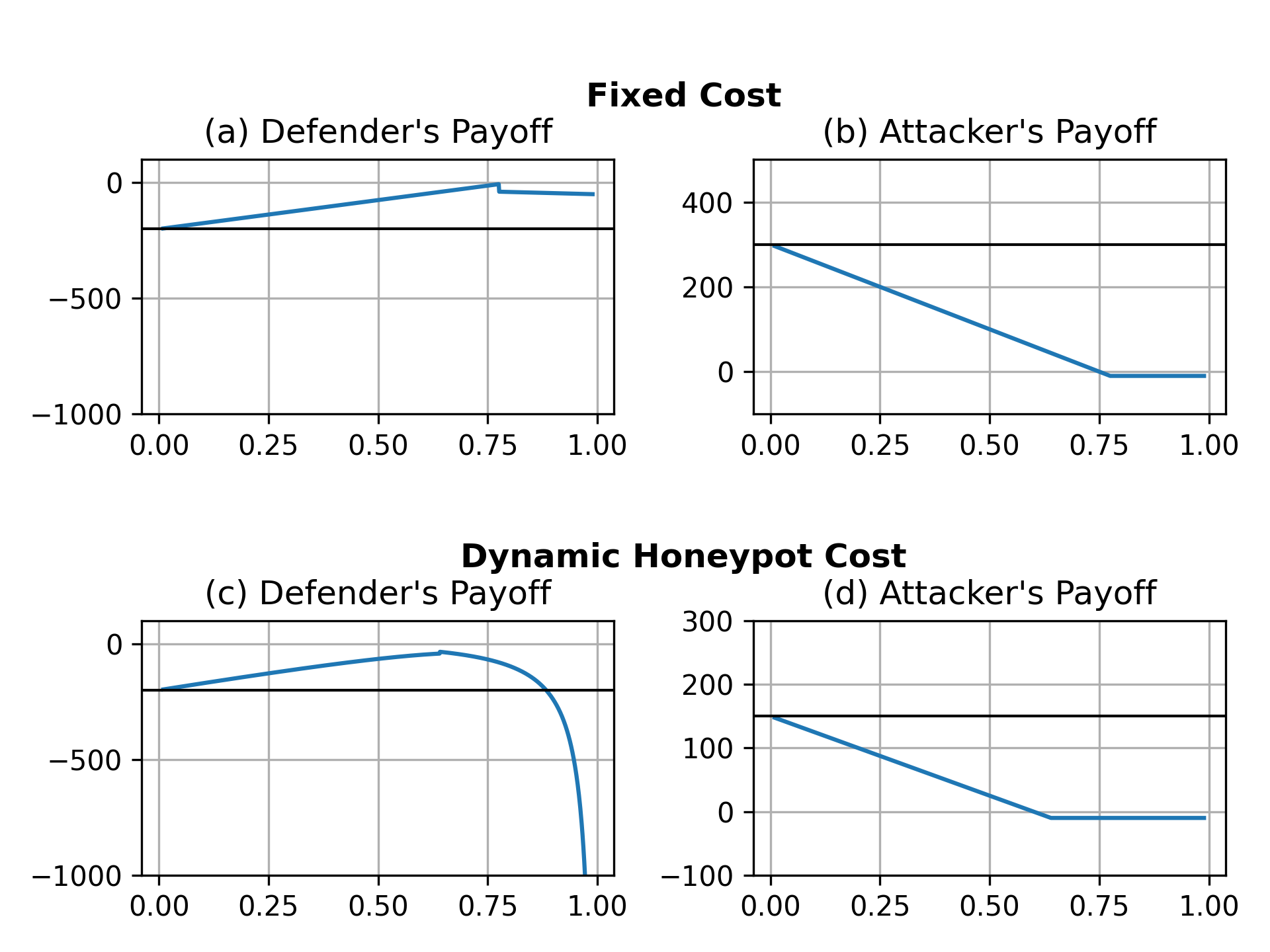}
        \centering
        \caption{Honeypot Deception Case Study}
        \label{fig: case study}
    \end{figure}

\section{Case Study} \label{section: case study}

In this section, we study two honeypot cases and show the payoff in which the defender and the attacker get in the honeypot detection game. In the first case, $c_h$ is given to the defender in a fixed state. In the second case, the honeypot node deployment cost $c_h$ varies with the ratio ($P_h$) given to the defender. We remark that all equilibrium payoffs identified in the above section are equal to the cost in Table \ref{table: payoff of equilibria}, except for the semi-separating equilibrium (FE3). Due to the dependence of the semi-separating equilibrium on extremely particular beliefs and the prior distribution of honeypots, it is challenging to meet the conditions in the real world. Therefore, this semi-separating equilibrium is excluded from the case study.

    \begin{table}[!tb]
    \centering
    \caption{Payoff of Pure Strategy Equilibria}
    \label{table: payoff of equilibria}
    \begin{tabular}{ccc}
    \Xhline{2\arrayrulewidth}
    \multirow{2}{*}{} & \multicolumn{2}{c}{Expected payoff} \T\B \\ \cline{2-3} 
                      & Defender         & Attacker         \T\B \\ \hline
    HE1               & $-c_a$           & $b_s$      \T\B   \\ \hline
    SE1               & $P_h(b_d - c_h) + (1-P_h)(-c_a)$    & $P_h(c_t) + (1-P_h)(b_s)$      \T\B   \\ \hline
    SE2               & $P_h(-c_h)$      & $-c_p$      \T\B   \\ \hline
    FE1               & $P_h(b_d - c_h) + (1-P_h)(-c_a)$    & $P_h(c_t) + (1-P_h)(b_s)$      \T\B   \\ \hline
    FE2               & $P_h(-c_h)$             & $-c_p$      \T\B   \\  \Xhline{2\arrayrulewidth}

    \end{tabular}
    \end{table}

    \subsection{Fixed honeypot cost experiment}
    
    In the first case, we only alter $P_h$, the probability that a honeypot node will be present in an environment where all other game variables are held constant, and examine the payoffs. In other words, the cost of operating a a specific strategy can be understood as the cost resulting from the honeypot nodes to deploy. The experimental settings are identical to those listed in Table \ref{table: experiment parameter}.

    Figure \ref{fig: case study} (a) -- (b) show the equilibria SE1/FE1 until a particular time. At this moment, as the honeypot probability, $P_h$, increases, the payoff of the defender gradually increases, and the payoff of the defender gradually declines. However, beginning at an interval $P_h \geq 0.7$ where the equilibrium strategy shifts, the attacker opts for the avoidance strategy. The attacker's payment remains constant at -10 at this moment. Even if the defender increases the frequency of honeypot exposure, the expense of controlling it climbs, resulting in a marginal reduction in payoff.

\begin{table}[!tb]
\centering
\caption{Parameters in Case Study}
\label{table: experiment parameter}
\begin{tabular}{ccx{2cm}x{2cm}}
\Xhline{2\arrayrulewidth}
                          & Parameter & \begin{tabular}[c]{@{}c@{}}Fixed \\ honeypot cost \end{tabular} & \begin{tabular}[c]{@{}c@{}}Dynamic \\honeypot cost \end{tabular} \\ \hline
\multirow{3}{*}{Attacker} & $b_s$        & 200          & 200                                                         \\
                          & $c_t$        & 100          & 100                                                         \\
                          & $c_p$        & 10           & 10                                                          \\ \hline
                          &              &              &                                                             \\ \hline
\multirow{4}{*}{Defender} & $b_d$        & 100          & 100                                                         \\
                          & $c_a$        & 300          & 300                                                         \\
                          & $c_h$        & 50           & $10 \cdot P_h / (1 - P_h)$                                                        \\
                          & $c_n$        & 30           & 30                                                          \\ \Xhline{2\arrayrulewidth}
\end{tabular}
\end{table}

    \subsection{Dynamic honeypot cost experiment}
    
    In the second case, we change the payoff of the defender as we mentioned in Section \ref{section: optimal decision}. With the ratio of honeypot nodes is $P_h$, the number of honeypot nodes to normal nodes is $P_h / (1 - P_h)$. The deception cost $c_n$ of the normal node is constant because the number of normal nodes is fixed, and only the honeypot node deployment cost $c_h$ changes in proportion to $P_h / (1 - P_h)$. The experimental parameters are shown below in the dynamic honeypot cost of Table \ref{table: experiment parameter}.

    Figure \ref{fig: case study} (c) -- (d) show the equilibria SE1/FE1 until a particular time. The point at which the equilibrium is converted is not affected by the honeypot deployment cost. Consequently, it is identical to the prior situation. However, when the frequency of honeypot exposure rises, i.e., as the number of deployed honeypots increases, the cost skyrockets and the defender's reward reduces precipitously after a given period of time. When $P_h$ is greater or equal to 0.85, the reward is less than -200, which is the case without the honeypot. Honeypot deployment cost has no effect on the attacker's payout, which is identical to the preceding scenario.

\section{Discussions} \label{section: discussions}

In this section, we discuss implication, limitations of our research and directions for future research.

\subsection{Comparison with Previous Works}
\label{subsec: comparison}

Among the previous research, Diamantoulakis et al. \cite{diamantoulakis2020game} is the most close to ours in that it explores the circumstance in which honeypot nodes are deployed into a system. Our study focuses on adding new honeypot nodes to the network, whereas their work focuses on the process of transforming some system resources into honeypot nodes in a system. Therefore, if the resources of the existing operating system are redundant, their method may be applied. However, if the operating system's current resources are insufficient to offer new services, honeypot nodes must be introduced through the allocation of new resources so that our study can be utilized effectively. In addition, our research classified the adoption of honeypot technology into three distinct phases, which were then analyzed and contrasted using game theory. We expect that the adoption of three tiers of honeypot technology will incur a variety of technical expenses. Our research offers the framework for system administrators to determine the economically and strategically suitable level of honeypot technology to use.

\subsection{Absence of Test Action}

This study does not offer any methodologies for determining whether or not a node discovered by an attacker is a normal node or a honeypot node. This is a limitation of the study. For the purposes of our study, we assumed that an adversary is unable to determine the true function of a node. Nevertheless, it possesses the technology to identify the target system of the attacker, such as assessing whether it is a virtual environment. 

As prior research has examined testing as an attacker's action, it is significant to implement the options considered by attackers in our extensive research. Nonetheless, in order to provide the operators with a decision-making aid that is unambiguous and useful, the set of operations should not be prepared in too many distinct ways, if at all possible. In addition, even if the attacker is able to conduct tests, we anticipate that the collection of PBE defense tactics will not shift significantly.

\subsection{Future Directions}
We propose directions for follow-up studies in three main directions.
The first is to compose one decision model with these three games as a subgame. As mentioned in Section \ref{subsec: comparison}, one of the significant implications of this study is that each model for introducing honeypot technology divided into stages was constructed. Future studies need to fully show the decision-making process of a system or network administrator by constructing an integrated game model with one more decision level that constitutes this model as a subgame.
The second is an analysis of the aspect of the game according to the attacker's ability to detect honeypots or related techniques. The attacker's ability to detect honeypots has not been considered much in the current model. The optimal ratio of honeypot nodes may vary depending on the type of attacker, so it is necessary to consider this in future studies.
The third is the analysis of the game according to the configuration of the repeated game. The current study consists of one-shot games. It needs to be shown that attackers and defenders achieve predictable equilibrium through interactive experiments.

\section{Conclusions} \label{section: conclusions}

Honeypot technology has been utilized for a considerable time and is a crucial framework for mitigating immediate or future hazards. Because defenders must convince attackers to target honeypots, honeypot operations need extensive strategic study.
We analyzed the honeypot deception games by applying the deception technique with new resource. In doing this, firstly we confirmed that an increase in deception action does not warrant an increase in the defender's payoff. Second, we demonstrated that the increase in the number of honeypot nodes does not always increase payoff but decreases after a certain maximum point. Furthermore, the honeypot selection cost can be dynamically increased, indicating that an increase in the honeypot node can drastically reduce the defender's payoff.
We studied how to operate the strategy in terms of defenders. Our analysis and results can help defenders make optimal decisions when applying the honeypot technique to their system. However, due to the lack of analysis of the repeated game, our work can be applied on a limited basis to memoryless attackers or environments where multiple attacks are difficult to perform. Further analysis of a comprehensive model and the repeated game should be analyzed in an extended study.

\section*{Acknowledgment}

This work was supported by Institute of Information \& communications Technology Planning \& Evaluation (IITP) grant funded by the Korea government(MSIT) (No.2021-0-00613, Zero Trust technology based access control and abnormal event analysis technology development for enterprise network protection in the untact era)

\bibliographystyle{IEEEtran}
\bibliography{references}

\end{document}